\documentclass[pra,showpacs,twocolumn]{revtex4-1}

\usepackage{amsmath}
\usepackage{graphicx}
\usepackage{bm}

\newcommand{\psiup}{\psi_\uparrow(x)}

\newcommand{\psiupdg}{\psi_\uparrow^\dagger(x)}

\begin{document}

\title{Quench-induced delocalization}
\author{Elmer V. H. Doggen}
\email[Corresponding author: ]{elmer.doggen@aalto.fi}
\author{Jami J. Kinnunen}
\affiliation{COMP Centre of Excellence and Department of Applied Physics, Aalto University, FI-00076 Aalto, Finland}

\begin{abstract}
We consider the evolution of an initially localized wave packet after a sudden change in the Hamiltonian, i.e.\ a quench.
When both bound and scattering eigenstates exist in the post-quench Hamiltonian, one might expect partial delocalization of the wave packet to ensue.
Here we show that if the quench consists of a sudden switching-off of short-range inter-particle interactions, then Tan's universal relations guarantee delocalization through the high-momentum tail of the momentum distribution.
Furthermore, we consider the influence of the range of the interaction and show how a finite range alters the coupling to highly excited states.
We illustrate our results using numerical simulations of externally trapped particles in one dimension.
If the external potential is both disordered and correlated, then the interaction quench leads to transport via delocalized states, showing that localization in disordered systems is sensitive to non-adiabatic changes in the interactions between particles.
\end{abstract}

\maketitle

\section{Introduction}

The non-equilibrium dynamics of quantum systems after suddenly changing a parameter in the Hamiltonian describing the system (a quantum quench) has been the topic of intense investigation recently, and is relevant for fundamental research into the thermalization of (closed) quantum systems, localization phenomena, chaos, decoherence and quantum information.
Such a quench can be realized in various ways using ultracold atoms, for instance by suddenly changing the shape of the trapping potential or by tuning the strength of the interactions between particles using a Feshbach resonance \cite{Kokkelmans2014a}.
A non-exhaustive list of recent experimental \cite{Pezze2004a, Kinoshita2006a, Strohmaier2007a, Schneider2012a, Brantut2012a, Gring2012a, Meinert2013a, Ronzheimer2013a, Langen2013a} and theoretical works \cite{Rigol2007a, Rigol2008a, Kollath2007a, Barthel2008a, Moeckel2008a, Eckstein2008a, Uhrig2009a, Iucci2009a, Moeckel2010a, Muth2010a, Kajala2011a, Canovi2012a, Mitra2012a, Iyer2012a, Pons2012a, Vidmar2013a, Mossel2012a, Caux2012a, Caux2013a, Mussardo2013a, DeNardis2014a, Bernier2014a, Rigol2012a, Rigol2014a} (for a review, see Ref.\ \cite{Polkovnikov2011a}) highlights the recent activity in this field.

In this work, we will consider the following scenario.
Suppose we have an interacting, localized and dilute cloud of particles, in the ground state of a Hamiltonian $\mathcal{H} = \mathcal{H}_0 + g\mathcal{H}_\text{int}$, where $\mathcal{H}_\text{int}$ represents the part of the Hamiltonian describing interactions between particles, and its magnitude can be tuned using the parameter $g$.
At a time $t = 0$ we now suddenly switch off the interactions, so that $g = 0$, and we follow the time evolution of the system.
Does the time evolution of this generic system have any universal properties?
We will show that this is indeed the case, provided that the interactions between particles are of short range.
For such a system, Tan \cite{Tan2008a,Tan2008b,Tan2008c} showed that there exist a number of universal relations describing the system.
One of these relations involves the high-momentum tail of the momentum distribution $n_q$, where $q$ is momentum:
\begin{equation}
  n_q \sim C/q^4, \,\,\,\, (q \rightarrow \infty). \label{contact}
 \end{equation}
The quantity $C$ is called the \emph{contact parameter} (see earlier related work \cite{Olshanii2003a,Gangardt2003a}) and can be interpreted as a measure for the probability of finding two particles in close proximity \cite{Zwerger2012a}.
The contact is also of interest because it can be measured experimentally \cite{Stewart2010a, Kuhnle2010a}, even dynamically \cite{Bardon2014a}.

If the non-interacting Hamiltonian $\mathcal{H}_0$ permits both bound and scattering (delocalized) states, then Eq.\ (\ref{contact}) implies that the particles, previously in the interacting ground state of the full Hamiltonian $\mathcal{H}$, must have some occupation of the scattering states after the interaction quench, which leads to (at least partial) delocalization.
We will demonstrate this principle, which we call \emph{quench-induced delocalization}, numerically for initially interacting particles in a one-dimensional system.
As a paradigmatic example of a system of which the single-particle eigenstates can be bound as well as delocalized, we first consider the simple case of a finite well, and analyze this case rigorously.
This system is also simple enough to permit investigating the influence of the range of inter-particle interactions.

We then proceed to consider a disordered potential, a case of particular interest to the study of Anderson localization (AL).
AL is the localization of single-particle eigenstates of a tight-binding Hubbard model with on-site disorder (see Sect.\ \ref{sec_disorder}), as described by Anderson in a seminal paper explaining the absence of diffusion in lattice systems with disorder \cite{Anderson1958a}.
This localization more generally affects the transport of classical waves as well as quantum particles through various media.
For instance, AL was experimentally reported for light waves \cite{Wiersma1997a, Scheffold1999a}, sound \cite{Hu2008a} and ultracold atoms \cite{Billy2008a, Roati2008a, Kondov2011a, Jendrzejeweski2012a}.
The particles are ``localized'' in the following sense: each eigenstate of the system is centered about some point in space $\bm{r}_0$ and has exponentially decaying tails, where the strength of the decay is determined by the \emph{Anderson localization length}.
The study of inter-particle interactions in disordered systems has received much interest recently \cite{SanchezPalencia2010a, Modugno2010a, Shapiro2012a, Fishman2012a, Bodyfelt2011a, Altman2004a, Lugan2007a, Lugan2007b, Fontanesi2009a, Flach2009a, Larcher2009a, Larcher2012a, Vosk2012a, Ristivojevic2012a, Ristivojevic2014a, Skokos2013a}.
In particular, it has been experimentally reported \cite{Lucioni2011a} that interactions destroy localization in disordered systems.
We will not study the interacting dynamics in this work, but instead show that as long as a transition or crossover to delocalized states exists in the single-particle spectrum, then localization will be sensitive to the switching-off of inter-particle interactions.

We will focus on the case where $N$ fermionic particles interact with a single impurity, where $N$ is either one (in which case the problem reduces to the case of two distinguishable particles) or up to a few dozen particles.
The case of two ($N=1$) interacting particles (TIP) can be considered a precursor of many-body effects.
As such, the TIP problem has been considered for instance in the context of many-body tunneling \cite{Rontani2012a, Rontani2013a} and AL \cite{Shepelyansky1994a,Krimer2011a}.
Recent experimental progress has made it feasible to study interactions between particles in few-body systems using ultracold gases \cite{Wenz2013a, Zurn2013a, Cheinet2008a, Ospelkaus2006a}.

This paper is structured as follows.
In Sect.\ \ref{sec_quench} we will outline the physical mechanism behind the quench-driven transport.
We then proceed, using the methods outlined in Sect.\ \ref{sec_methods}, by investigating specific systems as an example.
In Sect.\ \ref{sec_finitewell} we consider the simple case where the external trap is a square well of finite depth, which allows a detailed investigation of the properties of quench-induced delocalization, as well as the investigation of the range of the interaction.
Next, we discuss how the quench-induced delocalization mechanism applies to disordered systems in Sect.\ \ref{sec_disorder}.
We summarize and conclude in Sect.\ \ref{sec_conclusions}.

\section{Quench-induced delocalization} \label{sec_quench}

In this section, we will briefly discuss the general mechanism leading to delocalization after a non-adiabatic interaction quench.
We then proceed to discuss some examples using numerical simulations in one dimension.

Consider some interacting particles (fermions, bosons or a mixture thereof) in $d$ dimensions, trapped in an external potential $V_\text{ext}(\bm{r})$.
We will assume that the external potential permits at least some scattering (delocalized) single-particle eigenstates.
Note that many models commonly used in theoretical physics use potentials that do not have this property, but realistic physical potentials do.
For instance, Anderson's lattice model for disordered systems in one dimension does not permit scattering states (see Sect.\ \ref{sec_disorder}).
However, this property depends on the lack of correlations in the external potential.
A realistic potential will have short-range correlations on at least \emph{some} scale, with a corresponding crossover or transition to delocalized states in momentum space.

Now we assume that we have an initially localized wave packet, in the ground state of the full Hamiltonian $\mathcal{H} = \mathcal{H}_0 + g\mathcal{H}_\text{int}$, where $\mathcal{H}_0$ describes the non-interacting dynamics of the system.
The initial localization implies that the system will remain localized for all time if $\mathcal{H}$ is unchanged.
The parameter $g$ measures the strength of the interactions between the particles, described by the interacting part of the Hamiltonian $\mathcal{H}_\text{int}$.
Furthermore, we assume that the interactions have finite range $\bm{r}_0$.
For simplicity, we will set $\bm{r}_0 = 0$, although it is sufficient that $\bm{r}_0$ is small relative to the other length scales in the problem (in Sect.\ \ref{sec_finitewell} we will further investigate the influence of the range).
Under these conditions, it is known that the momentum distribution $n_q$ obeys Eq.\ (\ref{contact}).
In other words, the scattering states have some non-zero (only algebraically decaying with respect to momentum $q$) occupation probability.
This does not necessarily entail delocalization, because destructive interference between the components of the wave packet in scattering states may exist, so that only virtual particles exist in scattering states.
Such interference effects can be removed by introducing some source of dephasing to the system.
Here we will consider a possible source of dephasing: a non-adiabatic quench of the interaction parameter $g$ from some finite value to zero.
That is, at a time $t=0$ we switch off the interacting part of the Hamiltonian $\mathcal{H}_\text{int}$ and investigate the non-interacting dynamics for $t>0$.
We will show that such an interaction quench indeed leads to dephasing in the systems that we consider.
The argument holds for any $|g| > 0$, under the same (fairly minimal) conditions where the universal relation (\ref{contact}) is valid \cite{Zwerger2012a}.
The dephasing will lead to the real occupation of the scattering states, and therefore (at least partial) delocalization.

\section{Methods} \label{sec_methods}

We consider interacting particles in one dimension, as described by the following Hamiltonian:
\begin{align}
  & \mathcal{H} =  \sum_{\sigma}  \int dx  \, \psi_{\sigma}^\dagger(x)\Big[ -\frac{\hbar^2}{2m_\sigma} \frac{d^2}{dx^2} + V_\text{ext}(x) \Big] \psi_\sigma(x) \nonumber \\
  &+  \iint dx dx' \, \psiupdg \psi_\downarrow^\dagger(x') V_\text{int}(|x-x'|) \psi_\downarrow(x') \psiup, \label{mainhamiltonian}
\end{align}
where $x$ is position,  $m_\sigma$ is the mass of a particle, $V_\text{int}$ is the inter-particle potential, $\psi_{\sigma}^{(\dagger)}(x)$ destroys (creates) a particle of the kind $\sigma \in \{\uparrow, \downarrow\}$ and the external potential is given by $V_\text{ext}(x)$.
First, we consider the case where the external potential is a finite well (Sect.\ \ref{sec_finitewell}):
\begin{equation}
V(x) = \left\{
  \begin{array}{lr}
    -V_0 & \text{if} \,\,\,\, |x| \leq \Delta X/2,\\
    0 & \text{otherwise}. \label{eq_finitewell}
  \end{array}
  \right.
\end{equation}
Furthermore, we investigate the case where the external potential is given by either correlated or uncorrelated disorder (Sect.\ \ref{sec_disorder}):
\begin{align}
 V_{\text{uncorrelated},x} = & \mathcal{W}_x, \label{uncorr_dis} \\
 V_{\text{correlated},x} =  & \frac{1}{\sqrt{\pi\xi_\text{c}}} \sum_y \, \mathcal{W}_y \exp\Big[\frac{-(x-y)^2}{2\xi_\text{c}^2}\Big]. \label{corr_dis}
\end{align}
Here the uncorrelated disorder, in a discretized system, takes a random value on site $x$ in the interval $\mathcal{W}_x \in [-W/2,W/2]$ and $\xi_\text{c}$ is the correlation length (in units of the grid spacing).

\subsection{Two-particle case (numerically exact)}
We will describe the two-particle case in the following; the generalization to the variational approach used for $N+1$ particles is described below.
We solve the ground state of the Hamiltonian (\ref{mainhamiltonian}) for fixed parameters by minimizing $\langle \Psi | \mathcal{H} | \Psi \rangle$, where the exact wavefunction $|\Psi \rangle$ is given by:
\begin{equation}
 |\Psi \rangle = \sum_{mn} \phi_{mn} c_{\uparrow m}^\dagger c_{\downarrow n}^\dagger |0\rangle. \label{eq_twoparticleansatz}
\end{equation}
Here $c_{\sigma m}^\dagger$ creates a particle of the kind $\sigma$ in single-particle eigenstate $m$ ($m=0$ is the ground state of the non-interacting system) and $|0\rangle$ represents the vacuum.
The numerically exact procedure consists of minimizing $\langle \mathcal{H} \rangle$ with respect to the coefficients $\phi_{mn}$ using a method described in earlier work \cite{Doggen2014a}, thus obtaining the interacting ground state.

At a time $t = 0$, we suddenly switch off the interactions, so that $g = 0$.
The time evolution of the density (in units where $\hbar = 2m = 1$) is now given by:
\begin{equation}
 \langle \psi^\dagger_\uparrow(x,t) \psi_\uparrow(x,t) \rangle = \sum_{mnj} \phi^*_{mn} \phi_{jn} \alpha_j^*(x) \alpha_m(x) e^{i(E_j - E_m)t}, \label{density}
\end{equation}
where $\alpha_j$ denotes the $j$th single-particle eigenstate with energy $E_j$.
After a long time, we assume that the different energy states will have dephased.
This means the interference terms $m \neq j$ can be neglected, so that the time-averaged number density $\eta(x) = \lim_{t \rightarrow \infty} \overline{\langle \psi^\dagger_\uparrow(x,t) \psi_\uparrow(x,t) \rangle}$ (cf.\ Refs.\ \cite{Srednicki1994a,Rigol2012a,Mussardo2013a}) is given by:
\begin{equation}
 \eta(x) = \sum_{mn} |\phi_{mn}|^2 |\alpha_m(x)|^2. \label{eta}
\end{equation}
In the language of e.g.\ Refs.\ \cite{Rigol2008a, Caux2012a, Rigol2014a}, Eq.\ (\ref{eta}) is the ``diagonal ensemble'' (with respect to a specific observable: the density).
This ensemble was previously introduced in the present context by Deutsch \cite{Deutsch1991a}.
We will show in the following that the assumption of dephasing leading to Eq.\ (\ref{eta}) is justified \textit{a posteriori} for the systems we consider.

\subsection{$N+1$-particle case (approximative)}
For the $N+1$-particle system, we use the variational method based on Chevy's approach \cite{Chevy2006a} as described in Ref.\ \cite{Doggen2014a}.
In this case, we write the wavefunction approximately according to the ansatz:
\begin{equation}
 |\Psi \rangle = \sum_{mkn} \phi_{mkn} c_{\uparrow m}^\dagger c_{\uparrow k} c_{\downarrow n}^\dagger |FS\rangle,
\end{equation}
where $|FS\rangle$ represents the non-interacting Fermi sea with $N$ $\uparrow$-fermions in the lowest $N$ states.
This variational wavefunction includes all possible excitations with at most a single particle-hole excitation (it is possible to generalize the ansatz to multiple particle-hole excitations \cite{Giraud2009a}, but we will not consider such an extension here).
If $N=1$, then only one particle-hole excitation is possible, and the ansatz becomes exact and equivalent to Eq.\ (\ref{eq_twoparticleansatz}), as long as a procedure to find the variational coefficients can be found.
Analogous to the previous subsection we now find the expectation value of the density operator of the impurity $\downarrow$
\begin{equation}
 \langle \psi^\dagger_\downarrow(x,t) \psi_\downarrow(x,t) \rangle = \sum_{mknj} \phi^*_{mkn} \phi_{mkj} \alpha_n^*(x) \alpha_j(x) e^{i(E_n - E_j)t}, \label{eq_densityN}
\end{equation}
and the diagonal ensemble
\begin{equation}
 \eta_\downarrow(x) = \sum_{mkn} |\phi_{mkn}|^2 |\alpha_n(x)|^2.
\end{equation}

\section{Finite well} \label{sec_finitewell}

In the following section, we will discuss a simple system that permits both bound and scattering single-particle eigenstates of the non-interacting Hamiltonian $\mathcal{H}_0$, namely: two interacting particles in a finite well external potential given by Eq.\ (\ref{eq_finitewell}).
This system provides an intuitive illustration of quench-induced delocalization, and its simplicity allows us to characterize the process in great detail.
For instance, since the potential is relatively well-behaved, so are our numerics, and we can consider a wide range of initial interactions, including strongly repulsive interactions.
Also, since we do not require an ensemble average, we can investigate the numerically much more demanding case of finite-range interactions with a numerically exact method.
Another feature of this system is that it contains both bound and scattering single-particle states, a requirement for quench-induced localization.
We can then use the depth of the well to tune how many bound states are permitted in the system.

There is in fact a solution for TIP in a finite well using the Bethe ansatz \cite{Li1996a}.
Furthermore, there are solutions available for TIP in an infinite well \cite{Oelkers2006a} as well as a periodic potential \cite{Wouters2006a}, an impurity potential \cite{Zhang2013a}, two \cite{Busch1998a} and several \cite{Zinner2009a} particles in a harmonic trap and an approximate solution for a general external trap \cite{Ovchinnikov2004a} (for a recent review on few-body physics, see Ref.\ \cite{Blume2012a}).
Also, there are solutions for two bosons in the case of a double well \cite{Zollner2008a,Hunn2013a} and a delta function potential barrier \cite{SYKim2011a}.
However, the evaluation of physical observables such as the density from the Bethe ansatz solution is not straightforward.
Here we solve the TIP problem numerically, giving direct access to the wavefunctions and their time evolution.
This allows us to characterize the properties of the TIP problem in detail.
Note that the problems of two bosons and two distinguishable particles are equivalent, apart from a factor of $2$ in the interaction term of the Hamiltonian.

\begin{figure}[!hbt]
 \centering
 \includegraphics[width=\columnwidth]{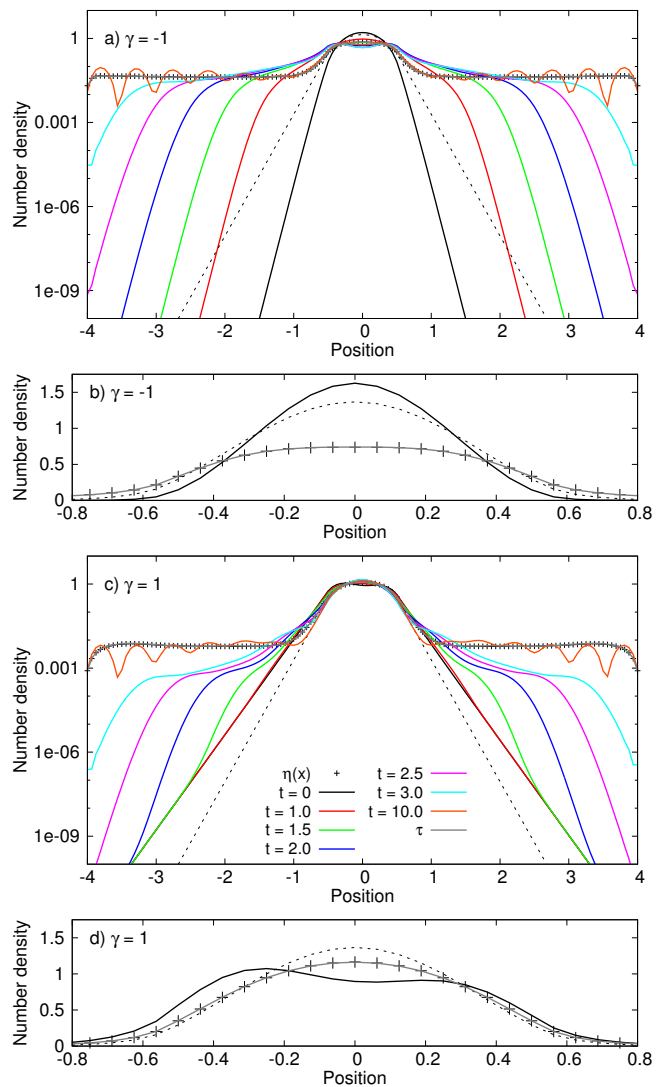}
 \caption{Number density (note the log scale) of an $\uparrow$-particle in a finite well initially interacting through repulsive or attractive contact interactions with a $\downarrow$-particle, a time $t$ after a quench to $\gamma = 0$, calculated numerically using Eq.\ (\ref{density}). Results shown are for an initial $\gamma = -1$ (attractive, panel \textbf{(a)}) and $\gamma = 1$ (repulsive, panel \textbf{(c)}). Panels \textbf{(b)} (attractive) and \textbf{(d)} (repulsive) show zoomed regions. Plus symbols show $\eta(x)$ calculated numerically using Eq.\ (\ref{eta}). The result given by $\tau$ -- the gray line coinciding with $\eta(x)$ -- is the result of averaging $51$ evenly spaced runs in the interval $t \in [100:200]$. The black dashed line shows the number density of the single-particle ground state. Note the effect of reflection from the boundary for $t = 10$. In panel \textbf{(d)}, a broken symmetry appears (see main text). The simulations use the parameters $L = 8, V_0 = 30$.} \label{timeevol}
\end{figure}

Let us first consider a potential of fixed depth $V_0$ and some fixed initial interaction between two particles of identical mass (see Fig.\ \ref{timeevol}).
We use a discretized system of length $L = 8\Delta X$ and we choose units where the well width $\Delta X = 1$, using the boundary condition that the wavefunction vanishes at the boundaries of the system ($x = \pm 4\Delta X$) and a discretized grid with 16 grid points per $\Delta X$.
Let us define a dimensionless interaction parameter $\gamma = g/V_0 \Delta X$, and check whether Eq.\ (\ref{eta}) is indeed reproduced for sufficiently long timescales for $\gamma = \pm 1$ (we choose $V_0 = 30$ and express time in units of $1/V_0$).
Note that the parameter $\gamma$ is not universal; different values of $g$ and $V_0$ with a constant $\gamma$ may give different results, as we will show below. 
At $t = 0$, the density (\ref{density}) decays exponentially, where the decay is stronger (weaker) for attractive (repulsive) interactions \cite{Zurn2013a} since the interaction energy causes the barrier to be effectively higher (lower).
This means that while the interaction \textit{does} couple to scattering states, destructive interference between the scattering states results in the localization of the interacting ground state.
Since we can explicitly track the time evolution of the system \emph{exactly}, we can verify to what extent Eq.\ (\ref{eta}), which assumes dephasing, is reproduced.
The agreement between the long-time average of the expectation value of the density operator (\ref{density}) and the diagonal ensemble (\ref{eta}) is excellent, as shown in Fig.\ \ref{timeevol}.
After the quench, waves emanate from the trap and start moving outwards light-cone-like \cite{Cheneau2012a,Carleo2014a} with a wavefront velocity approximately equal to $V_0 \Delta X/\hbar$ for the values of $\gamma$ we have considered.
This is consistent with the existence of a Lieb-Robinson bound (a similar bound is found in the disordered case of Sect.\ \ref{sec_disorder}); although the system is non-interacting after the quench, the deterministic time evolution of the system implies that correlations in the initial interacting state are preserved.
The waves reflect from the boundary of the system and move back and forth indefinitely (in an open system, the waves would move outward to infinity).

\subsection{The long-time limit}

\begin{figure}[!htb]
 \centering
 \includegraphics[width=\columnwidth]{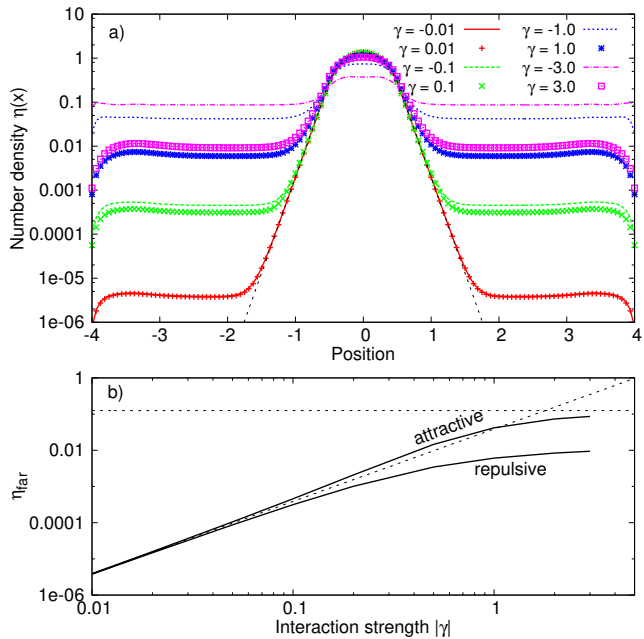}
 \caption{(\textbf{a}) Number density (note the log scale) of an $\uparrow$-particle in a finite well initially interacting through repulsive (symbols) or attractive (lines) contact interactions with a $\downarrow$-particle, a long time after an interaction quench to $\gamma = 0$. Results shown are for an initial $\gamma = \pm 0.01, \pm 0.1, \pm 1.0, \pm 3.0$. The black dashed line shows the number density of the single-particle ground state. $L = 8, V_0 = 30$. (\textbf{b}) Asymptotic density tail value $\eta_\text{far}$ (note the double log scale) evaluated at $x = 2.5$ as a function of the interaction strength $\gamma$, for both repulsive (lower solid line, $\gamma >0$) and attractive (upper solid line, $\gamma < 0$) interactions. The sloped dashed line is given by $\kappa \gamma^2/L$ ($\kappa = 0.31$), the horizontal dashed line is equal to $1/L$.} \label{2partdens}
\end{figure}

Let us henceforth focus on the function $\eta(x)$, the density of a particle in the long-time limit after an interaction quench.
Fig.\ \ref{2partdens}a shows the value of $\eta(x)$ for various values of $\gamma$ and fixed $V_0 = 30$.
Outside the well $\eta(x)$ approaches a constant value (let us define this value as $\eta_\text{far}$) rather than continuing the exponential decay of the single-particle or interacting ground state.
In the weakly interacting limit ($|\gamma| \ll 1$) $\eta_\text{far}$ is independent of the sign of $\gamma$ and scales $\propto \gamma^2$, cf.\ Ref.\ \cite{Doggen2014a} (see Fig.\ \ref{2partdens}b).
By contrast, for stronger interactions the attractive case ($\gamma < 0$) results in a \emph{larger} value of $\eta_\text{far}$ compared to the repulsive case ($\gamma > 0$).
Although this may seem counter-intuitive (the decay of the interacting ground state is \emph{stronger} in the attractive case), it can be understood in terms of the momentum distribution (\ref{contact}).
Attractive particles are more likely to be found at the same position, resulting in a larger value of the contact parameter \cite{Barth2011a,Doggen2013a}, and therefore the coupling to scattering states is stronger.
For moderately attractive interactions, $\eta_\text{far}$ increases more rapidly than $\propto \gamma^2$, which then crosses over when $\gamma \approx 1$ to a regime where the increase is slower than $\propto \gamma^2$.
In the limit that $\gamma \rightarrow -\infty$, $\eta(x)$ is expected to approach the constant value $1/L$ (shown in Fig.\ \ref{2partdens}b) as the probability of occupying single-particle bound states becomes negligible compared to the occupation probability of scattering states.

For strongly repulsive interactions, $\eta_\text{far}$ looks to be saturating to a fixed value, in accordance with the saturation of the contact parameter \cite{Barth2011a,Doggen2013a}.
The sloped dashed line in Fig.\ \ref{2partdens}b shows the weakly interacting limit $\eta_\text{far} = \kappa \gamma^2/L$.
$\kappa$ is a dimensionless constant; its value can be inferred from a simple perturbative calculation.
In the weakly interacting limit, the occupation probability of a scattering state with momentum $q$ and energy $E_q$ is approximately $p_q = (gn_0)^2/(2E_q - 2E_0)^2$, where $n_0$ is the density \cite{Kinnunen2012a}.
The total number of particles in scattering states divided by $L$ is then (in the continuum limit) $1/(2\pi L) \int dq \, p_q = (\gamma^2/L) \kappa$.
If we approximate $n_0 \approx 1/\Delta X$ and $E_0 \approx -V_0$, we obtain $\kappa \approx 0.31$, which is the same value as obtained using a fit.
We have verified that the value of $L\eta_\text{far}$ is independent of the system size.
Thus, the constant value of $\eta_\text{far}$ for fixed $L$, while an artifact of the finiteness of the system and the boundary conditions, should be understood as a measure of the fraction of particles occupying scattering states, independent from the boundary conditions.
Also, our results converge with respect to the spacing of the grid (we use 16 grid points per $\Delta X$).
As a check on the robustness of our result, we repeated various simulations using an inverted Gaussian potential and found no qualitative differences.
Furthermore, the ground state energy obtained for a harmonic trap using our method agrees with the exact result of Busch et al.\ \cite{Busch1998a}.

\begin{figure}[!htb]
 \centering
 \includegraphics[width=\columnwidth]{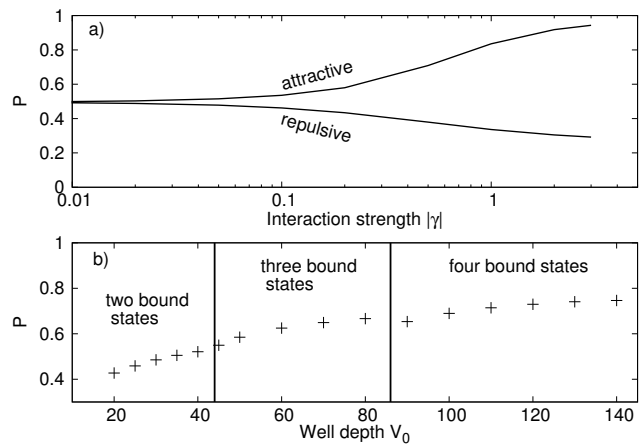}
 \caption{(\textbf{a}) Occupation probability of excited states $|\phi_{nn}|^2$ as a function of the energy of the single-particle state $E_n - E_0$. Symbols show the repulsive case $\gamma = 1$, the solid line shows the attractive case $\gamma = -1$. The dashed line is a guide to the eye and has a $1/E^2$ decay. Results shown use 64 grid points per $\Delta X$ and an energy cutoff to reduce the aliasing error. (\textbf{b}) Quasiparticle weight (plus symbols, left y-axis) and $|\phi_{01}|^2$ (crosses, right y-axis) as a function of $\gamma$ ($V_0 = 30$).} \label{oddeven}
\end{figure}

Since Eq.\ (\ref{eta}) describes the late-time properties of the system, all of the relevant physics is contained within the occupation numbers $|\phi_{mn}|^2$.
In Fig.\ \ref{oddeven}a we show the contact tail ($\propto 1/q^4$), which manifests itself in the elements $|\phi_{nn}|^2$.
Since the high-energy scattering states are almost plane waves with a dispersion $\propto q^2$ (cf.\ Ref.\ \cite{Doggen2014a}), this tail exhibits a $1/E^2$ decay, where $E$ is the energy (expressed in units of $\hbar^2/2m\Delta X^2$).
Fig.\ \ref{oddeven}b shows the elements $|\phi_{00}|^2$ and $|\phi_{01}|^2$ as a function of the interaction strength $\gamma$.
The former can be identified with the quasiparticle weight, which is equal to $1$ for zero interactions and is reduced for stronger interactions.
For weak interactions, there is an odd-even effect in the occupation numbers, so that $\phi_{mn} = 0$ if $|m - n|$ is odd.
This effect is due to the even symmetry of the problem, which is broken at finite interaction, as is visible in the density profile of Fig.\ \ref{timeevol}d.
The total density of both particles remains symmetric since $|\phi_{mn}| = |\phi_{nm}|$.
On the repulsive side, there is a sharp transition around $\gamma \approx 0.7$, which depends only weakly on $g$ for different values of $V_0$; for instance, at $V_0 = 30$ the transition is around $g \approx 21$ and at $V_0 = 100$ it is approximately $g \approx 26$.
This is because the symmetry breaking is related to the difference between the first two energy levels in the trap.
However, for sufficiently deep wells this is independent of $V_0$.
Conversely, the effect is much weaker on the attractive side.
The breaking of symmetry for repulsive interactions can be associated with the onset of Luttinger liquid behavior and spin-charge separation.

\subsection{Pair correlations} \label{sec_paircorr}

\begin{figure}[!htb]
 \centering
 \includegraphics[width=\columnwidth]{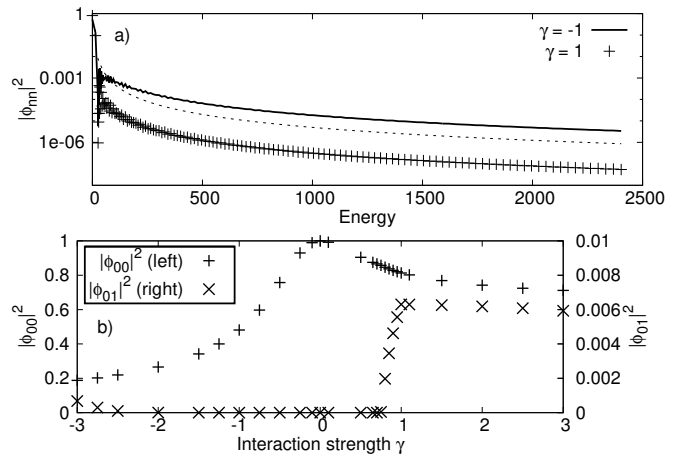}
 \caption{(\textbf{a}) Conditional probability $P$ that an $\uparrow$-particle is found in a scattering state, given that the $\downarrow$-particle is in a scattering state, as a function of $\gamma$ (note the log scale). Both repulsive (lower line, $\gamma > 0$) and attractive (upper line, $\gamma < 0$) interactions are considered. Values computed using Eq.\ (\ref{cond_prob_eq}). $L = 8, V_0 = 30$. (\textbf{b}) $P$ as a function of $V_0$ for fixed interaction $\gamma = -0.01$. Vertical lines separate regions where the number of single-particle bound states is constant; due to finite size-effects, the transition is not sharp.} \label{cond_prob}
\end{figure}

To characterize transport of particles away from the well, we consider the conditional probability:
\begin{equation}
 P = \frac{P(\uparrow, \downarrow \text{in scattering states})}{P(\downarrow \text{in scattering state})} = \frac{\sum_{m_sn_s} |\phi_{m_sn_s}|^2}{\sum_{mn_s} |\phi_{mn_s}|^2}, \label{cond_prob_eq}
\end{equation}
where the summation over $m_s$ ($n_s$) is restricted to single-particle scattering states and the summation over $m$ runs over all states.
This probability $P \in [0,1]$ can thus be interpreted as the probability of finding a $\uparrow$-particle in a scattering state, given that the $\downarrow$-particle is in a scattering state.
In Fig.\ \ref{cond_prob}a we show this value as a function of $\gamma$ for fixed $V_0$.
Interestingly, for weak interactions ($\gamma \ll 1$) there is a substantial conditional probability of finding two particles in a scattering state, even though the probability of finding the first particle in a scattering state decays as $\gamma^2$.
$P$ further increases and approaches $1$ for strongly attractive interactions, suggesting a role of pair correlations \cite{Zurn2013a}.
Meanwhile, $P$ is suppressed for strongly repulsive interactions.
Fig.\ \ref{cond_prob}b shows the dependence of $P$ on $V_0$ in the weakly interacting limit ($\gamma = -0.01$).
This dependence has the peculiar feature that the conditional probability of finding a particle in a bound state decreases as the depth of the well is increased, even though the number of bound states as well as the energy difference between bound and scattering states increase, suggesting that the pair tunneling effect is enhanced for deeper wells.
Note, however, that while the \emph{conditional} probability increases, the probability of finding at least one particle in a scattering state after the quench does decrease for deeper wells, as expected.
Numerical constraints limit the range of values we can consider for $V_0$, prohibiting a full quantitative analysis of the influence of the well depth.
 
\subsection{Finite-range interactions}

The problem of interacting particles with a finite-range interaction potential is considerably more difficult to handle than contact interactions, especially in a system without translation invariance (e.g., an externally trapped system).
Fortunately, assuming zero-range interactions is often a good approximation for realistic physical systems, especially in one dimension, where the $\delta$-function potential does not require renormalization.
Nevertheless, it is of interest to investigate the influence of the range of the inter-particle potential \cite{Valiente2010a, Lee2012a, Lode2012b, Hohenadler2012a, Streltsov2013a, Schachenmayer2013a}.

In this work we will model the finite range of the inter-particle potential by a Gaussian with standard deviation $r_0$ (in units of $\Delta X$), and cut off the potential at $2r_0$.
We normalize the Gaussian so that the integral over real space is equal to $g$.
Therefore, in the limit that $r_0 \rightarrow 0$, we recover the $\delta$-function potential with strength $g$.
Otherwise, we use the same parameters as in the preceding sections; i.e.\ the well depth $V_0 = 30$, the system size $L = 8$ and the well width $\Delta X/L = 1/8$.
Energies are expressed in units of $\hbar^2/2m\Delta X^2$.

\begin{figure}[!htb]
 \centering
 \includegraphics[width=\columnwidth]{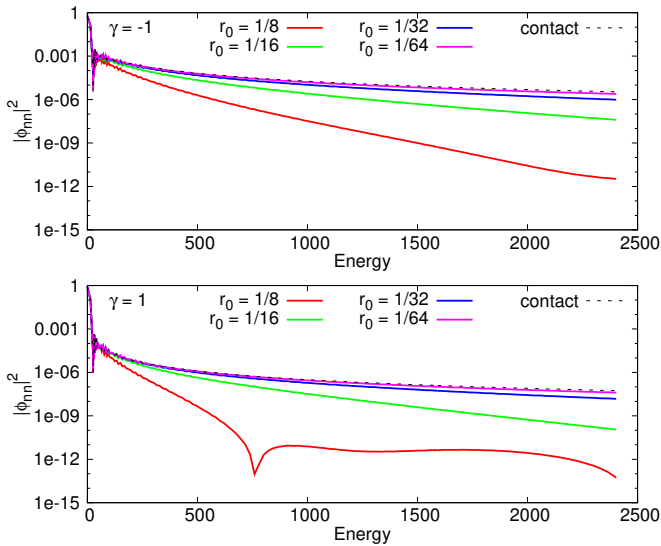}
 \caption{Occupation probability of excited states $|\phi_{nn}|^2$ as a function of the energy of the single-particle state $E_n - E_0$, for attractive ($\gamma = -1$, top) and repulsive ($\gamma = 1$, bottom) interactions. As the range of the interaction increases, the deviation from the $1/E^2$ scaling expected from Eq.\ (\ref{contact}) increases. The dashed line indicates the result obtained by using zero-range contact interactions.} \label{fig_finrange}
\end{figure}

We consider the influence of the range $r_0$ on the contact tail.
The result is plotted in Fig.\ \ref{fig_finrange}, which can be compared to Fig.\ \ref{oddeven}a.
As the range $r_0$ is increased, the deviation with respect to the universal scaling increases, and non-universal behavior appears as the energy approaches the energy $\hbar^2/2mr_0^2$ associated with the range; the occupation numbers then show behavior that depends on the details of the inter-particle potential (such non-universal features are clearly visible in the result for $r_0 = 1/8$ and $\gamma = 1$, see Fig.\ \ref{fig_finrange}).
Recently, this breakdown of the contact regime has been experimentally reported in nuclear matter, where $r_0^{-1}$ differs from the Fermi momentum by less than an order of magnitude and universal behavior only appears within a limited momentum window \cite{Hen2014a}.
Theoretically, this crossover to a regime where the finite range of the interaction is significant was investigated from the viewpoint of the high-frequency tail in radio-frequency spectroscopy \cite{Braaten2010a}.
Note that the effect of the finite range is generally to induce a reduction in the coupling to highly excited states, in agreement with the aforementioned experimental and theoretical results.
This reduction will be significant when the energy of the lowest scattering state is comparable to or higher than the energy associated with the range of the inter-particle interactions.

\section{Disordered potential} \label{sec_disorder}

\subsection{Anderson localization}

In Anderson's original work \cite{Anderson1958a} he considers a system described by the following Hamiltonian (the Anderson-Hubbard Hamiltonian):
\begin{align}
 \mathcal{H}_\text{Anderson} = & -  J \sum_{i\sigma} c_{i\sigma}^\dagger c_{i+1\sigma} + \text{H.c.} \nonumber \\
 &  + U \sum_i c_{i\uparrow}^\dagger c_{i\uparrow} c_{i \downarrow}^\dagger c_{i \downarrow} + \sum_{i\sigma} V_\text{ext}^i  c_{i\sigma}^\dagger c_{i\sigma}, \label{andersonmodel}
\end{align}
where $c_{i\sigma}^{(\dagger)}$ destroys (creates) a particle of the kind $\sigma \in \{\uparrow,\downarrow\}$ on lattice site $i$, $J$ is the hopping parameter, which determines the kinetic energy required for particles to tunnel to adjacent lattice sites, and $U$ determines the strength of inter-particle interactions (in Anderson's paper, $U = 0$).
The external (on-site) potential $V_\text{ext}^i$ is given by Eq.\ (\ref{uncorr_dis}), with a different random value at each site.
Anderson showed that a remarkable property of this Hamiltonian is that \emph{all} eigenstates of the system, regardless of their energy, are localized in the sense that the large-distance density tail of any eigenstate centered about some point in space $x_0$ decays at least as fast as $e^{-2x'/\xi_\text{A}}$, where $x' = |x-x_0|$ is the distance from the initial point and $\xi_\text{A}$ is the \emph{Anderson localization length}.
In other words, the Anderson localization length $\xi_\text{A}$ determines the decay of the least localized single-particle eigenstate, which may be populated by an arbitrary wave packet.
The feature of exponential localization of all eigenstates actually holds only in one and two dimensions \cite{Abrahams1979a}.
In three dimensions, there is an energy cutoff, the \emph{mobility edge}, beyond which the states are no longer localized.
This transition from localized to extended states is captured approximately by the Ioffe-Regel criterion \cite{Ioffe1960a}.

However, the appearance of Anderson localization for all eigenstates in one dimension depends on the assumption that the disorder is uncorrelated from site to site.
In experiments with ultracold atoms, the disordered potential is often created using a speckle potential \cite{SanchezPalencia2010a}, where the correlation between two points of the potential decays over a characteristic length scale $\xi_\text{c}$, the correlation length.
For bosons in one dimension as described by the Gross-Pitaevskii equation, it has been shown \cite{SanchezPalencia2007a} that the correlation length induces an \emph{effective mobility edge}.
The speckle potential causes the localization length to increase strongly beyond this effective mobility edge  \cite{Gurevich2009a,Lugan2009a}.
Clearly, any physically realistic potential must have some short-range correlations, otherwise the Fourier spectrum of the potential is not bounded and the potential will have significantly populated frequency components corresponding to arbitrarily high energies.
Even in the case of a lattice, where the assumption of uncorrelated disorder from site to site might be justified, the interaction will couple to higher Bloch bands, which will resolve any correlation that exists on length scales shorter than the distance between lattice sites \cite{Doggen2014a}.
It now follows that, due to Eq. (\ref{contact}), an interaction quench of the type considered here will lead to partial delocalization through single-particle delocalized states.

\begin{figure}[!hbt]
 \centering
 \includegraphics[width=\columnwidth]{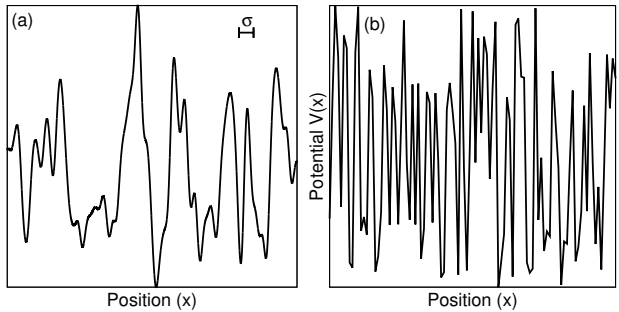}
 \caption{Schematic depiction of a Gaussian correlated \textbf{(a)} and an uncorrelated \textbf{(b)} potential. The correlation length $\xi_\text{c}$ is depicted in panel \textbf{(a)}.} \label{schematic_potential}
\end{figure}

Let us consider a disordered external potential, where the potential is given by Eqs.\ (\ref{uncorr_dis}-\ref{corr_dis}).
The potential may be uncorrelated from site to site, or have correlations between sites, schematically depicted in Fig.\ \ref{schematic_potential}.
We will here consider Gaussian correlations only, with a correlation length $\xi_\text{c}$.
This correlated potential is slightly different from the speckle potential used in many experiments \cite{SanchezPalencia2010a}, but the long-range properties of such a potential are difficult to take into account in a closed, finite system.
We do not expect major differences between this potential and the speckle potential such that our conclusions will be altered \cite{Kondov2011a}, although the dependence of the localization length $\xi_\text{A}$ on the energy is subtly different.
In the case of a speckle potential, the localization length can increase as a function of energy in finite energy windows \cite{Piraud2012a}, whereas in the case of Gaussian correlations $\xi_\text{A}$ diverges monotonically \cite{Izrailev2012a}.
Strictly speaking, there is no mobility edge in the case of Gaussian correlations, since the localization length diverges smoothly as a function of momentum \cite{Izrailev2012a}.
However, $\xi_\text{A} \rightarrow \infty$ as $k \rightarrow \infty$ as long as there is some kind of finite-range correlation, and there is a crossover to delocalized states in the case of Gaussian correlated disorder.
For the uncorrelated potential, we can use known values of $\xi_\text{A}$ \cite{Izrailev1998a} (uniquely determined by $W$) to benchmark our results.

We consider particles interacting through a contact potential $V_\text{int} = g\delta(x - x')$, so that the universal relation (\ref{contact}) is valid for all momenta $q \gg \mathcal{L}$, where $\mathcal{L}$ is any physically relevant length scale in the problem.
With this approximation, we solve the ground state of the Hamiltonian (\ref{mainhamiltonian}) in a discretized system with $L$ sites (we use units where the grid spacing $\delta x = 1$ and energies are expressed in terms of $\hbar^2/2m\delta x^2$), with the boundary condition that the wavefunction vanishes at the boundary.
For reliable numerical results, we would ideally like to have weak disorder $W \ll 1$, $\xi_\text{A} \ll L$, $\xi_\text{c} \ll L$ and $\xi_\text{c} \gg 1$.
We cannot hope to fulfill these conditions and still compute the interacting ground state without further approximations.
Therefore, we relax the requirement for weak disorder.
This means that the single-particle ground state will populate a state in the so-called Lifshits tail \cite{Lifshits1988a, Modugno2010a}, where it is `classically' trapped.
However, this is not a major issue, because the interactions lead to the population of most single-particle states, including the states in the Lifshits tail, the Anderson-localized states beyond the Lifshits tail as well as all delocalized states in the tail of the momentum distribution (\ref{contact}).
We also relax the requirement that $\xi_\text{c} \gg 1$ and instead choose $\xi_\text{c} = \mathcal{O}(1)$.
This means that the ratio of localized to delocalized states in the discrete system with correlated disorder is also of $\mathcal{O}(1)$, resulting in a relatively low population of delocalized states.
However, this does not affect the result qualitatively.

Eq.\ (\ref{eta}) implies that after the quench most of the single-particle eigenstates will have some occupation.
Therefore, the initial wave packet, which is strongly localized due to occupying the lowest state in the Lifshits tail (in the weakly interacting limit), will start to expand.
A long time after the quench, the long-distance density tail is then expected to be dominated by the least localized state in the single-particle eigenspectrum.
With uncorrelated disorder, this state decays according to the Anderson localization length $\xi_\text{A}$.
In the case of correlated disorder, the least localized state is actually delocalized, resulting in the delocalization of a certain fraction of the wave packet.
This delocalization depends on the initial interaction, since this determines the coupling to scattering states.
Specifically, since the contact scales as $g^2$ for $|g| \ll 1$ \cite{Barth2011a, Kinnunen2012a, Doggen2013a} we expect this fraction to be proportional to $g^2$ in the weakly interacting limit.

\subsection{Numerical results}

\begin{figure}[!hbt]
 \centering
 \includegraphics[width=\columnwidth]{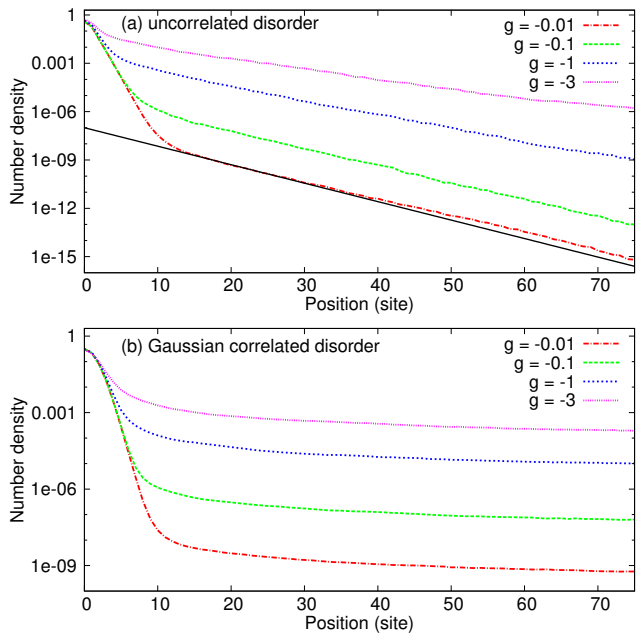}
 \caption{Number density (note the log scale) of a particle in a disordered potential, initially in the interacting ground state, interacting with a second particle with interaction strength $g$. The plotted density is obtained from Eq.\ (\ref{eta}) and shows the density a long time after an interaction quench to $g = 0$ (averaged over 250 runs). Panel \textbf{(a)} shows the result for uncorrelated disorder (\ref{uncorr_dis}) and panel \textbf{(b)} shows correlated disorder (\ref{corr_dis}). The solid line in panel \textbf{(a)} is a guide to the eye and decays according to the Anderson localization length obtained from Ref.\ \cite{Izrailev1998a}.} \label{fig_and1}
\end{figure}

We consider a system of size $L = 151$, either uncorrelated disorder or $\xi_\text{c} = 2$ and various values of the initial interaction strength $g$.
We choose a value of the disorder strength so that $\xi_\text{A} \ll L$, but we also want to use as weak a disorder as possible under this constraint so that the approximate solution for $\xi_\text{A}$ is valid and the numerics are well-behaved; we take $W = 4$, for which $\xi_\text{A} \approx 7.6$ \cite{Izrailev1998a}.
Because the peak density will take an almost random position somewhere on the grid, we only consider the tail in the direction where the grid has the most sites until the boundary of the system is reached.
The result is shown in Fig.\ \ref{fig_and1}, where we plot Eq.\ (\ref{eta}), the long-time density $\eta(x)$ for various values of $g$.
Since the peak density will be a minimum distance $L/2$ away from the edge of the system, we plot $\eta(x)$ as a function of distance away from the peak density in the interval $[0,75]$.
We take the average of 250 runs, using a different random seed for each run, where the average is obtained by logarithmic averaging.
Note that a similar approach was used to study a non-interacting wave packet in a disordered potential \cite{Moratti2012a}.
In our case, the interaction quench combined with the relation (\ref{contact}) guarantee that the \emph{phase randomization ansatz} discussed in Ref.\ \cite{Moratti2012a} is valid for an arbitrary initial wave packet with significant overlap of interacting particles (see also Ref.\ \cite{Piraud2011a}).
Indeed, it is valid even when starting from the interacting ground state, deep in the Lifshits tail.

In the case of uncorrelated disorder, the long-distance density tail reproduces the approximate prediction for the localization length well, as expected.
Interestingly, there is an enhancement of $\xi_\text{A}$ as the initial interactions become stronger (a fit of the slope of the exponential decay of the density yields $\xi_\text{A} = 14 \pm 1$ for $g = -3$), even though the particles are non-interacting after the quench.
The decay is thus weaker than the decay of the least localized eigenstate of the system.
A similar increase of the localization length has been predicted for interacting systems without an interaction quench \cite{Shepelyansky1994a,Krimer2011a}, although it should be noted that in our simulations, where the initial state has a large overlap between the two particles, the enhancement may be artificially large \cite{Krimer2011a}.

In the case of correlated disorder, localization is much weaker.
There is still a slight decay of the density, where it might be expected that $\eta(x)$ approaches a constant value, corresponding to delocalization (see Sect.\ \ref{sec_finitewell}).
This is an artifact of our numerical constraints; to reach this constant regime we should be in the regime where the momentum $q \gg 1/\xi_\text{c}$.
However, since $\xi_\text{c} = 2$ these high-energy states cannot be fully resolved.
This is also reflected in the least localized state of the single-particle eigenspectrum; while the localization length diverges exponentially for high-energy states \cite{Izrailev2012a}, the discreteness of the system implies an energy cutoff, which implies that the least localized state has a weak but finite exponential decay.
For both uncorrelated and correlated disorder, the strength of the density tail scales as $\propto g^2$ for weak interactions, as expected.
Note that in the weakly interacting limit $|g| \ll 1$, the result is independent of the sign of $g$ (cf.\ Sect.\ \ref{sec_finitewell}).
For strongly repulsive interactions $g \geq 1$ our iterative method to find the interacting ground state fails, possibly because of the near-degeneracy of low-energy single-particle states that are nevertheless separated in real space such that their spatial overlap is almost zero.

\begin{figure}[!hbt]
 \centering
 \includegraphics[width=\columnwidth]{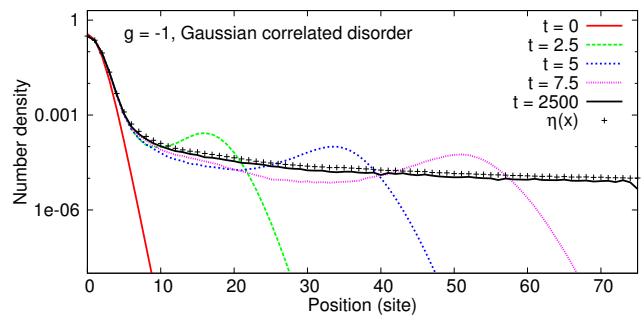}
 \caption{Number density (note the log scale) of a particle in a disordered potential, initially in the interacting ground state, interacting with a second particle with interaction strength $g$. The plotted density is obtained by computing the density a time $t$ after an interaction quench to $g = 0$ (averaged over 250 runs). For comparison, symbols show the result obtained from Eq.\ (\ref{eta}).} \label{fig_and2}
\end{figure}

What is the timescale associated with delocalization?
To answer this question, we compute the time evolution explicitly using Eq.\ (\ref{density}), see Fig.\ \ref{fig_and2}.
We express time in dimensionless units $t \rightarrow tW/\hbar$.
In these units, the ``switching on''-time associated with dephasing is $t \approx 1$, cf.\ Sect.\ \ref{sec_finitewell}.
There is, however, still a small deviation from the density at long times compared to Eq.\ (\ref{eta}), which was also found in Ref.\ \cite{Moratti2012a}.

\begin{figure}[!hbt]
 \centering
 \includegraphics[width=\columnwidth]{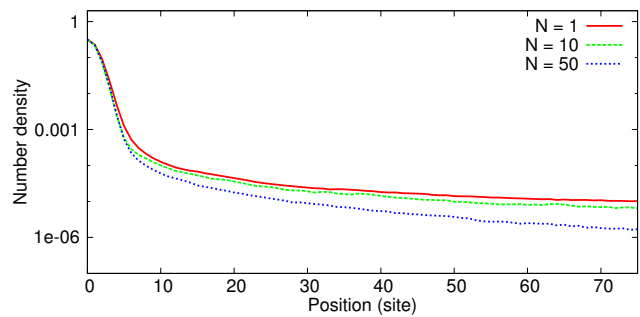}
 \caption{Number density (note the log scale) of a particle in a disordered potential, initially in the interacting ground state, interacting with $N$ fermionic particles with interaction strength $g = -1$, a long time after an interaction quench to $g = 0$. A Gaussian correlated disordered potential is used. Note that for the $N=10$ and $N=50$ runs we have averaged over 100 runs.} \label{fig_and3}
\end{figure}

It is also of interest to study the case where $N$ ``$\uparrow$''-fermions interact with a single ``$\downarrow$''-impurity.
For this purpose we use the variational approach of our earlier work \cite{Doggen2014a} to compute the long-time density $\eta_\downarrow(x)$ of the impurity.
Although the solution is not exact, we obtain qualitative agreement with the exact solution for the two-particle case, see Fig.\ \ref{fig_and3}.
One might expect the effect of the initial interactions to increase as the number of majority component particles $N$ is increased; this is certainly the case in the mean-field approximation, where the Hartree shift is proportional to the majority component density.
However, the variational calculation, for a moderate interaction strength $g = -1$, shows a decrease in the tail density.
The presence of additional majority component particles restricts the possible excitations of the majority component particles due to Pauli blocking.
Furthermore, since the first few excited single-particle states are likely to be, on average, strongly localized elsewhere in space, they have negligible overlap with the impurity in the ground state and do not contribute significantly to scattering events to highly excited states.

\section{Conclusions} \label{sec_conclusions}

In conclusion, we consider a localized, interacting system where the interactions are turned off instantaneously (quenched).
We show that as long as some scattering states exist and the inter-particle interactions have zero range, the quench leads to partial delocalization, and we characterize this delocalization quantitatively for certain specific systems.
Furthermore, we show how the range of the interactions influences the coupling to highly excited states, as expected from the high-momentum tail of the momentum distribution (\ref{contact}).
An interaction quench thus leads to transport through delocalized states in disordered systems, provided they exist.

The resulting transport might be observed in an experiment with ultracold atoms akin to Ref.\ \cite{Zurn2013a}, where the advantage of our proposed setup is that single-particle tunneling can be neglected.
Although we consider the one-dimensional case, we expect qualitatively similar effects in higher dimensions.
We assume in this work that the interaction quench is infinitely fast, whereas the dynamics associated with higher energy states also becomes increasingly fast as a function of energy.
Nevertheless, we expect that the delocalization will be dominated by the lowest scattering states, which have the highest occupation probability and with which the slowest dynamics is associated, so that a sufficiently fast sweep should be feasible.

We stress that while we study specific realizations of interacting systems, we expect the same mechanism to hold for a greater number of particles, since the momentum tail (\ref{contact}) holds in general for any number of particles.
Also, it is noteworthy that partial delocalization depends algebraically rather than exponentially on the size of the change in the interaction parameter (for small changes).
This means that for the quench-induced delocalization to be vanishingly small, the non-adiabatic change in the interactions itself must be vanishingly small.
Furthermore, the precise shape of the external potential is irrelevant, as long as some bound and scattering states exist, and the initial interacting wave packet is localized.
Indeed, if one considers a system that is repeatedly quenched, then our results suggest that eventually all particles will transfer to scattering states.
On the other hand, if the variation in the interaction strength is sufficiently small or gradual, or if the lowest-energy delocalized state is of sufficiently high energy, then delocalization of a significant number of particles through this process will occur on timescales much larger than laboratory timescales.

It would be of interest to consider the case where the interaction parameter is quenched from a fixed value to some other, arbitrary non-zero value.
Is the delocalization mechanism in this case the same, considering that one is quenching to a non-integrable system rather than to an integrable one?
It has been suggested that this is not the case \cite{Gogolin2011a, Serbyn2013a}, but is not possible to address this question with the method used in this work.
Nevertheless, if we consider a quench to a \emph{small} value of the interaction strength, i.e.\ a quench from $g$ to $g'$, where $g \gg g'$ and $g' \ll 1$, then the analysis of Yurovsky and Olshanii \cite{Yurovsky2011a} suggests that for sufficiently small timescales our analysis is still valid.

As a final remark, we stress that our results should not be used to draw conclusions about the long-standing problem of interacting particles in a random potential.
A common belief is that in these systems, the inter-particle interactions will drive transport.
However, this mechanism is distinct from the transport induced by non-adiabatic variations in the interactions between particles.

\emph{Acknowledgments.}--- This work was supported by the Academy of Finland through its Centres of Excellence Programme (2012-2017) under Project No.\ 251748.
We thank F.\ Massel, M.\ O.\ J.\ Heikkinen, A.\ Harju, Z.\ Fan, A.\ Uppstu, J.-P.\ Martikainen, P.\ T\"orm\"a, J.\ E.\ Baarsma and J.\ Suhonen for useful discussions and suggestions, and A.-M.\ Visuri for discussions and sharing unpublished results.
LAPACK \cite{Lapack1999} was used for our computations.

\bibliographystyle{apsrev4-1}
\bibliography{ref}

\end{document}